\documentclass[twocolumn]{aastex62}

\shorttitle{BP disc iron line reverberation}
\shortauthors{Wenda Zhang et al.}

\begin{document}
\newcommand{\swsixteen}{Swift J1644+57}
\newcommand{\swtwenty}{Swift J2058.4+0516}


\title{Probing the Bardeen-Petterson effect in tidal disruption events with spectral line reverberation mapping}
\correspondingauthor{Wenda Zhang and Wenfei Yu}
\email{zhang@asu.cas.cz, wenfei@shao.ac.cn}


\author[0000-0003-1702-4917]{Wenda Zhang}
\affiliation{Astronomical Institute, Czech Academy of Sciences, Bo\v cn\'i II, CZ-141 31 Prague, Czech Republic}

\author[0000-0002-3844-9677]{Wenfei Yu}
\affiliation{Shanghai Astronomical Observatory and Key Laboratory for Research in Galaxies and Cosmology,
Chinese Academy of Sciences, 80 Nandan Road, Shanghai 200030, China: wenfei@shao.ac.cn}

\author{Vladim\'ir Karas}
\affiliation{Astronomical Institute, Czech Academy of Sciences, Bo\v cn\'i II, CZ-141 31 Prague, Czech Republic}

\author{Michal Dov\v ciak}
\affiliation{Astronomical Institute, Czech Academy of Sciences, Bo\v cn\'i II, CZ-141 31 Prague, Czech Republic}





\begin{abstract}
For an inclined accretion flow around a rotating black hole, the combined effect of the
Lense-Thirring precession and viscous torque tends to align the inner part of the flow with the black hole spin,
leading to the formation of a warped disc, known as the Bardeen-Petterson effect \citep{bardeen_lense-thirring_1975}.
In tidal disruption events (TDEs) in which a super-massive black hole starts to accrete the bound debris, if the black hole is spinning, in general the stellar orbit is inclined with the black hole spin. So is the accretion disc formed following circularization and radiative cooling of the debris.
\citet{xiang-gruess_formation_2016} studied in detail the stellar debris evolution and disc formation in TDEs when the stellar orbit is
inclined, and found that a warped disc would form under certain conditions.
In this work we investigate properties of time-resolved
fluorescent iron line originating from a warped disc that is irradiated by the initial X-ray flare.
We find that the time-resolved spectrum shows distinct features before and after a critical time.
This critical time depends on the Bardeen-Petterson radius $r_{\rm BP}$, i.e., the outer boundary of the inner aligned disc; 
while the line width during the later stage of the X-ray flare is sensitive to the inclination of the outer disc flow.
This demonstrates that time-resolved X-ray spectroscopy can be a powerful tool to probe the Bardeen-Petterson effect in 
TDE flares and can be used to measure the Bardeen-Petterson radius as well as put constraint on
the black hole mass and spin. 
\end{abstract}




\section{Introduction}
Particles moving around rotating black holes are subject to Lense-Thirring precession 
if their orbit is inclined with the black hole symmetry axis \citep{lense_uber_1918}.
As found out by \citet{bardeen_lense-thirring_1975}, for an inclined accretion disc around a rotating black hole, 
the combined effect of the Lense-Thirring (LT) precession and viscous torque tends 
to align the inner part of the disc with the black hole
spin, resulting in a warped disc (hereafter BP disc). This is known as the Bardeen-Petterson (BP) effect.

Later studies found that the behavior of the disc under this effect is sensitive to the thickness and effective viscosity
of the disc
\citep{papaloizou_time-dependence_1983}.
For a thin disc ($H/R \leq \alpha$, where $H$, $R$, and
$\alpha$ are disc scale height, radius, and viscous parameter, respectively), the warp propagates diffusively; while 
for a thick disc the warp is propagated by bending wave \citep{papaloizou_dynamics_1995}.

In tidal disruption events (TDEs) the super-massive black hole tidally disrupts a nearby star that falls within the tidal radius and accretes part of the
stellar debris
\citep{rees_tidal_1988,evans_tidal_1989,phinney_manifestations_1989}. 
Approximately one half of the debris are located on bound orbits. 
The bound debris circularizes and cools radiatively, and subsequently forms an accretion disc \citep[e.g.,][]{evans_tidal_1989,kochanek_aftermath_1994,hayasaki_circularization_2016,
bonnerot_disc_2016}. 

If the supermassive black hole is spinning, in general we expect that the stellar orbit is not aligned with the black hole spin axis, 
hence the disc would be inclined as well. \citet{franchini_lense-thirring_2016} discussed the effect of 
the Lense-Thirring precession on TDEs.
\citet{xiang-gruess_formation_2016} studied in detail the hydrodynamic evolution of the stellar debris whose orbit
is inclined with the black hole spin, and the disc formation process following cooling of the debris.
They found that a misaligned disc could form in TDEs provided that the warp propagation time is large compared with 
the local accretion time and/or the natural alignment radius is small. \citet{liska_bardeen-petterson_2018} performed
GRMHD simulations to study the evolution of an misaligned disc around a spinning black hole, and found that a Bardeen-Petterson disc
could form rather soon ($\sim 200-500~\rm GM~c^{-3}$ after the accretion starts).

In \citet[][hereafter Paper I]{zhang_predictions_2015} we studied the time-resolved spectrum of 
the fluorescent iron K$\alpha$ line
originating from an aligned accretion disc in TDEs, and found that the iron line reverberation can be used to constrain 
properties of the black hole as well as the accretion flow.
In this paper we extend our work to most of the TDE cases in which a warped disc
is involved and investigate the response of the iron line profile originating from
the \textit{BP disc} due to the initial irradiation by the central TDE flare.

Although this paper is motivated to study the temporal-spectral features 
of the fluorescent iron line emission originating from a misaligned disc caused by the BP effect, 
the conclusion we draw is quite general and can be applied to warped discs induced by other mechanisms, such as 
radiation pressure induced warping \citep{pringle_self-induced_1996,pringle_self-induced_1997}.
But the irradiation we study here is not from a delta-function flare but a rising edge of a bright
TDE flare, which causes a full ionization of the disc flow after the ionization front passes by.  
While with current X-ray facilities it may be not easy to detect TDEs in the early rising phase,
the Square Kilometre Array (SKA), which will start scientific observations in the mid-2020s, may be able to detect the
the precursor of potential X-ray flaring of TDEs, e.g., radio flaring as mentioned in
\citet{yu_early_2015}, making X-ray observations of TDEs in the early rising phase possible.

\section{Model set-up}
We model the BP disc as two independent components, both of which rotate around the black hole with distinct inclination. Transition of the
two components occurs at the Bardeen-Petterson radius $r_{\rm BP}$. 
The inner disc has an observer's inclination $i_{\rm in}$, with the outer boundary of
$r_{\rm BP}$, and is truncated at the Innermost Stable Circular Orbit with radius $r_{\rm ISCO}$;
while the outer disc has an observer's inclination $i_{\rm out}$, with the  
outer boundary set at 1000 $\rm GM/c^2$, and the inner part is truncated at $r_{\rm BP}$. We neglect the
transition layer between the two disc components for simplicity. In calculating the spectrum of the outer disc component,
we treat it as if it is located on the
equatorial plane. Since the outer disc component is relatively far away from the black hole, we expect that the shift of line energy is dominated by
Doppler effect while the general relativistic effects are negligible.  In our calculation, the black hole has
a dimensionless spin
parameter $a\equiv Jc/GM^2$, where $J$ and $M$ are the black hole's angular momentum and mass, respectively;
$c$ is the speed of light, and $G$ is the gravitational constant.
We calculate the evolution of the iron line spectra for the two components separately using the 
\textsc{KYcode} \citep{dovciak_XSPEC_2004,dovciak_extended_2004} and then sum up the spectra.
For details of our calculation for a single disc component, the reader is kindly referred to Paper I.

In the calculations we set the fluorescent iron energy to be 6.4 keV, 
the rest frame energy of the neutral iron K$\alpha$ line. 
We assume that the emission of the iron line is isotropic in the
rest frame of the disc fluid. The radial dependence of iron line emissivity is parameterized by the emissivity index $q$, 
such that the emissivity in the disc fluid rest frame $\epsilon(r) \propto r^{-q}$. In the calculations we
take $q=3$, and in Section~\ref{sec:robustness} we will examine
the effect of our particular choice of the emissivity law on the results.
In fact, due to the diversity of the geometries and the stars disrupted in TDEs,
the emissivity would be quite different from what we expect in AGNs. 
%
%

\section{Results}
\subsection{An illustrative example}
\label{sec:example}
\begin{figure*}
 \includegraphics{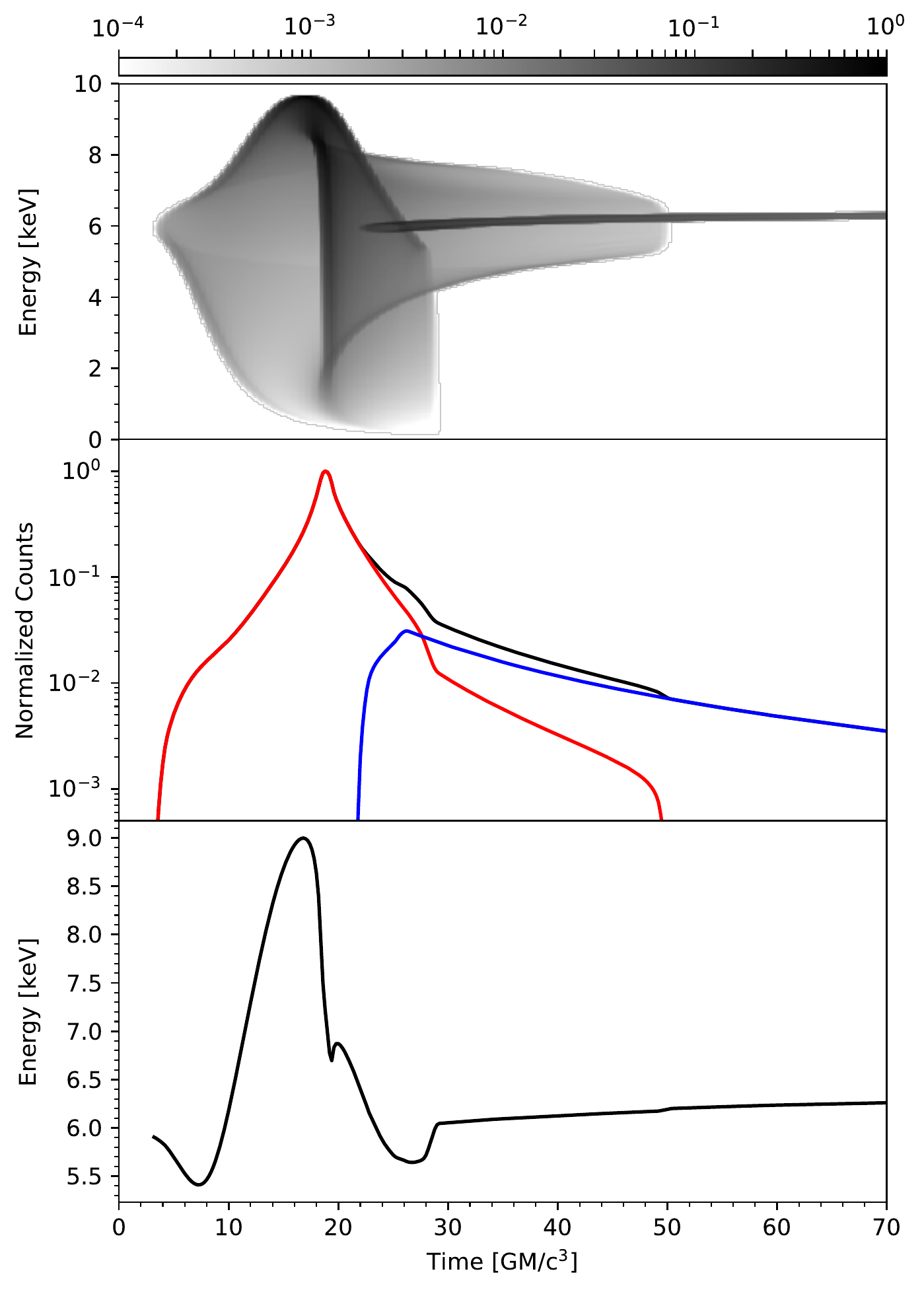}
 \caption{Top panel: the evolution of the fluorescent iron line emission originated from a BP disc.
 Parameters: black hole spin $a=0.9987$, inner disc inclination: $i_{\rm in} = 85^\circ$, 
 outer disc inclination: $5^\circ$, and truncation radius $r_{\rm BP} = 20~\rm GM/c^2$.
 Middle panel: lightcurve of the iron line emission. The contribution from inner and outer discs are plotted in
 red and blue solid lines, respectively; while the black solid line represents the sum of the two. 
 Bottom panel: time evolution of centroid iron line energy. For definition of the centroid line energy one is referred to Paper I.
 \label{fig:case0}}
\end{figure*}

In the top panel of Fig.~\ref{fig:case0} we show the time evolution of the iron line spectrum for a BP disc with an edge-on
inner disc ($i_{\rm in}=85^\circ$) and a face-on outer disc ($i_{\rm out}=5^\circ$), to highlight the distinct iron
line dynamic profile from the two disc components. The black hole is taken as maximally rotating ($a=0.9987$), and the transition
radius $r_{\rm BP}=20~\rm GM/c^2$. In the middle and bottom panels, we also plot the emission line light curve and the evolution of the centroid line energy,
respectively, following our practice in Paper I. We can see that before $t\sim50~\rm GM/c^3$, the iron line is highly broadened, and contains
a ``loop'' shape structure that is characteristic of the profile from an accretion disc with high observer's inclination.
As discussed in Paper I, the loop structure is caused by the truncation of the disc at ISCO due to the hole at the center, which carries information of the black hole spin $a$ and the inner disc inclination $i_{\rm in}$. 
\citet{reynolds_x-ray_1999} calculated the response of iron line emitted by an accretion in AGNs to the
variability of the primary hard X-ray. In the transfer functions, they found ``loop'' structure for high-inclination discs
and ``tail'' structure for low-inclination discs, similarly with what we found for 
TDE discs when the rising edge of the irradiation X-ray flare is considered (paper I).

The narrow iron line from the outer disc emerges at $t\sim22~\rm GM/c^3$ and dominates the emission line spectrum starting from
$t\sim 25 ~\rm GM/c^3$. The energy of the narrow component shifts blueward with time, from $\sim6$ keV to 6.4 keV,
the rest frame energy of the iron line. This ``tail'' shape is typical for low-inclination disc.
The contrast between temporal-spectral features before and after $50~\rm GM/c^3$ shows that 
iron line reverberation can be a powerful tool to detect the BP disc and measure the BP radius.

With the emission line profile at earlier stage of the TDE flare, we can put constraint on $a$ and $i_{\rm in}$ following Paper I. 
Out of the four key parameters $a$, $r_{\rm BP}$, $i_{\rm in}$, and $i_{\rm out}$, there are $i_{\rm out}$ and $r_{\rm BP}$ left to be constrained.
In the following we focus on the effects of varying $i_{\rm out}$ and $r_{\rm BP}$.

\subsection{Varying $i_{\rm out}$}
\begin{figure*}
 \includegraphics{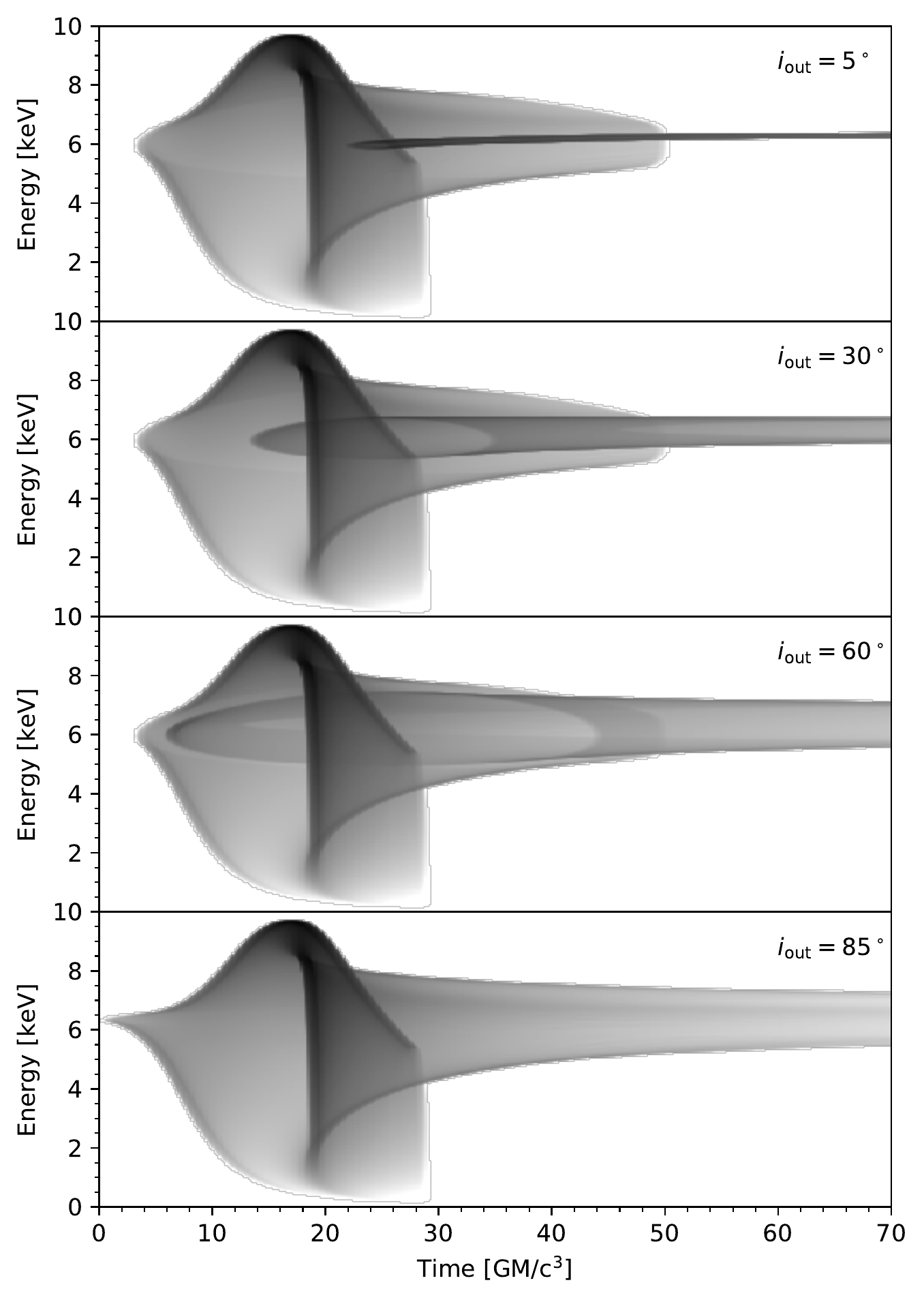}
 \caption{The evolution of fluorescent iron line spectrum from BP discs with
 an inner disc inclination of $85^\circ$ and various outer disc inclinations.
 Parameters: black hole spin $a=0.9987$, and the
 truncation radius $r_{\rm BP} = 20~\rm GM/c^2$. Outer disc inclinations: $5^\circ$, $30^\circ$,
 $60^\circ$, $85^\circ$ from top to bottom.
 \label{fig:incl}}
\end{figure*}

\begin{figure}
 \includegraphics[width=\columnwidth]{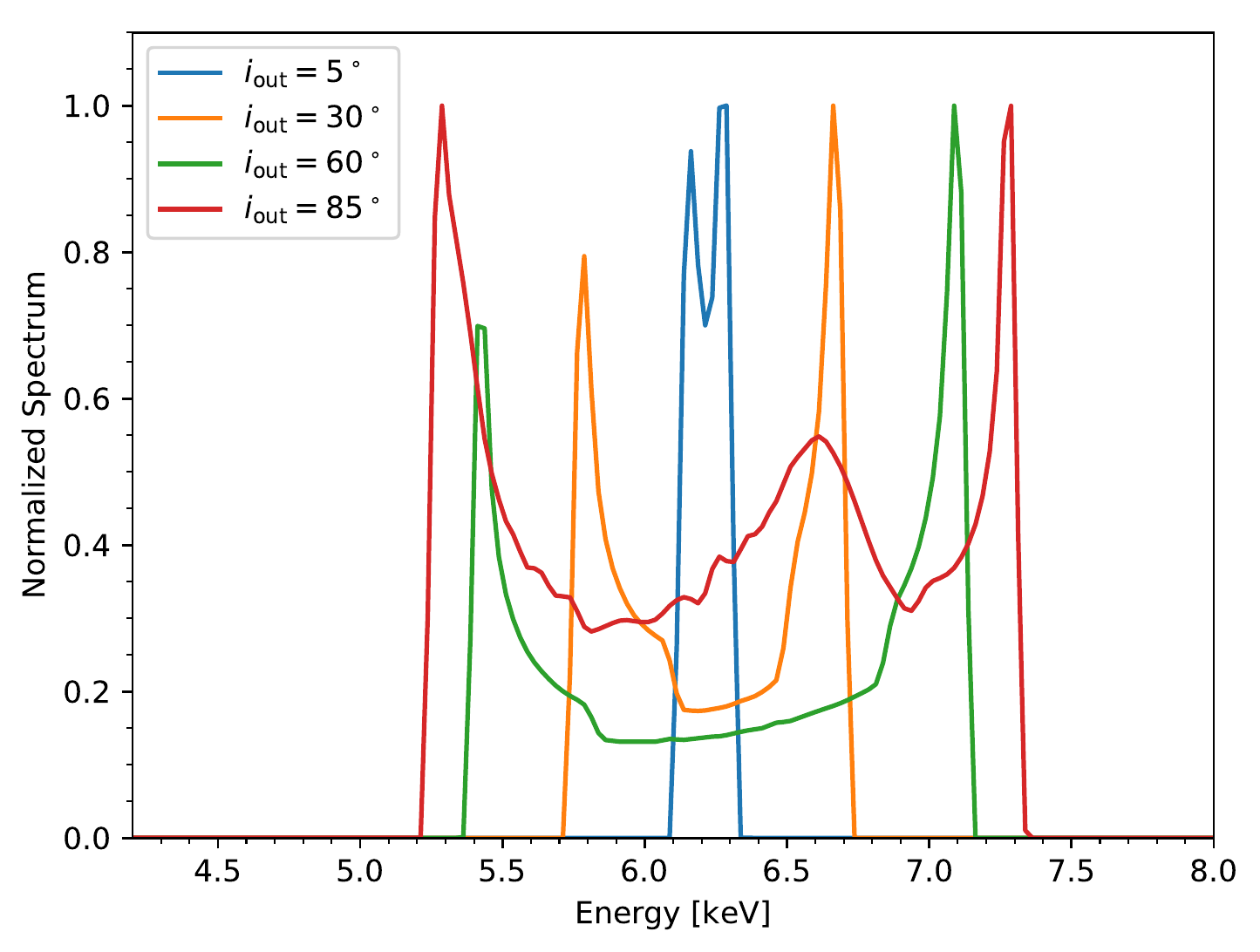}
 \caption{The iron line spectrum at $t=55~\rm GM/c^3$, corresponding to Fig.~\ref{fig:incl}.
 \label{fig:line}}
\end{figure}

We present the time-resolved spectra of various $i_{\rm out}$ in Fig.~\ref{fig:incl}. The inner disc inclination is
fixed to be $85^\circ$. After $t\sim50~\rm GM/c^3$ when the iron line contains contribution from the outer disc only,
the line width increases with $i_{\rm out}$. To inspect the line profile,  
In Fig.~\ref{fig:line} we plot iron line spectra at $t=55~\rm GM/c^3$.
All lines show double-horned profile, and the separation of the two horns increases with inclination. 
This is the character of emission line from Newtonian, Keplerian disc \citep[see, e.g.,][]{fabian_broad_2000}.
For the outer disc, the general relativistic
effects become insignificant and the Doppler effect dominates the broadening of the iron line. 

\begin{figure}
 \includegraphics[width=\columnwidth]{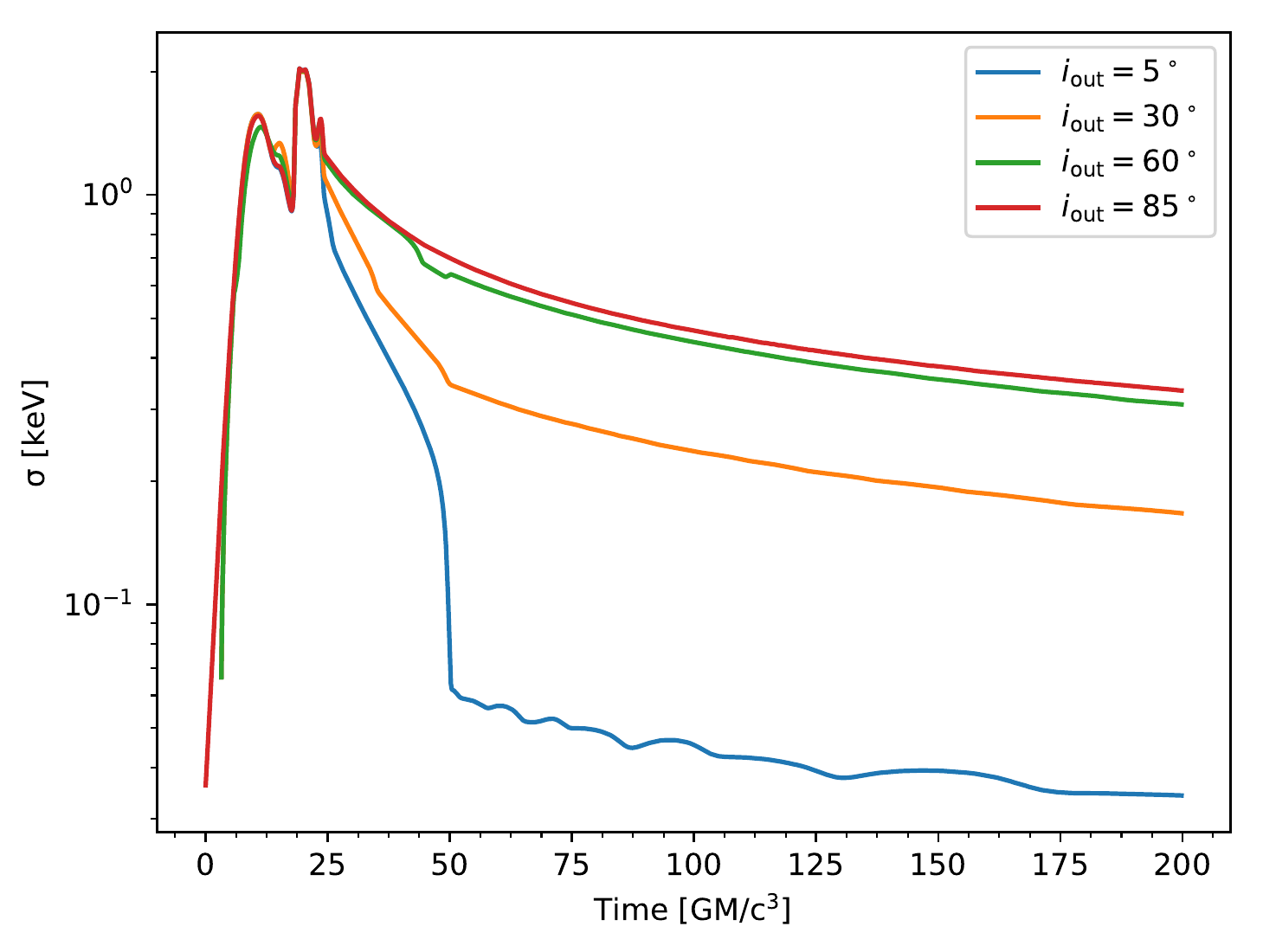}
 \caption{The evolution of the iron line standard deviation $\sigma$, corresponding to Fig.~\ref{fig:incl}.\label{fig:sigma}}
\end{figure}

To trace the evolution of the line profile, we calculate the standard deviation $\sigma$ 
of the line as a proxy of the horn separation, which is defined as 
\begin{equation}
  \sigma = \sqrt{\frac{\sum_i r_i(E_i - \bar{E})^2}{\sum_i r_i}},
\end{equation}
where $r_i$ is the iron line rate in the $i$-th energy bin, and 
\begin{equation}
\bar{E} \equiv \frac{\sum_i r_i E_i}{\sum_i r_i}
\end{equation}
is the iron line mean energy. In Fig.~\ref{fig:sigma} we present the evolution of $\sigma$ 
for different iron line profile corresponding to Fig.~\ref{fig:incl}. Increase of $\sigma$ with $i_{\rm out}$ is
clearly seen.


\subsection{Varying $r_{\rm BP}$}
\begin{figure*}
 \includegraphics{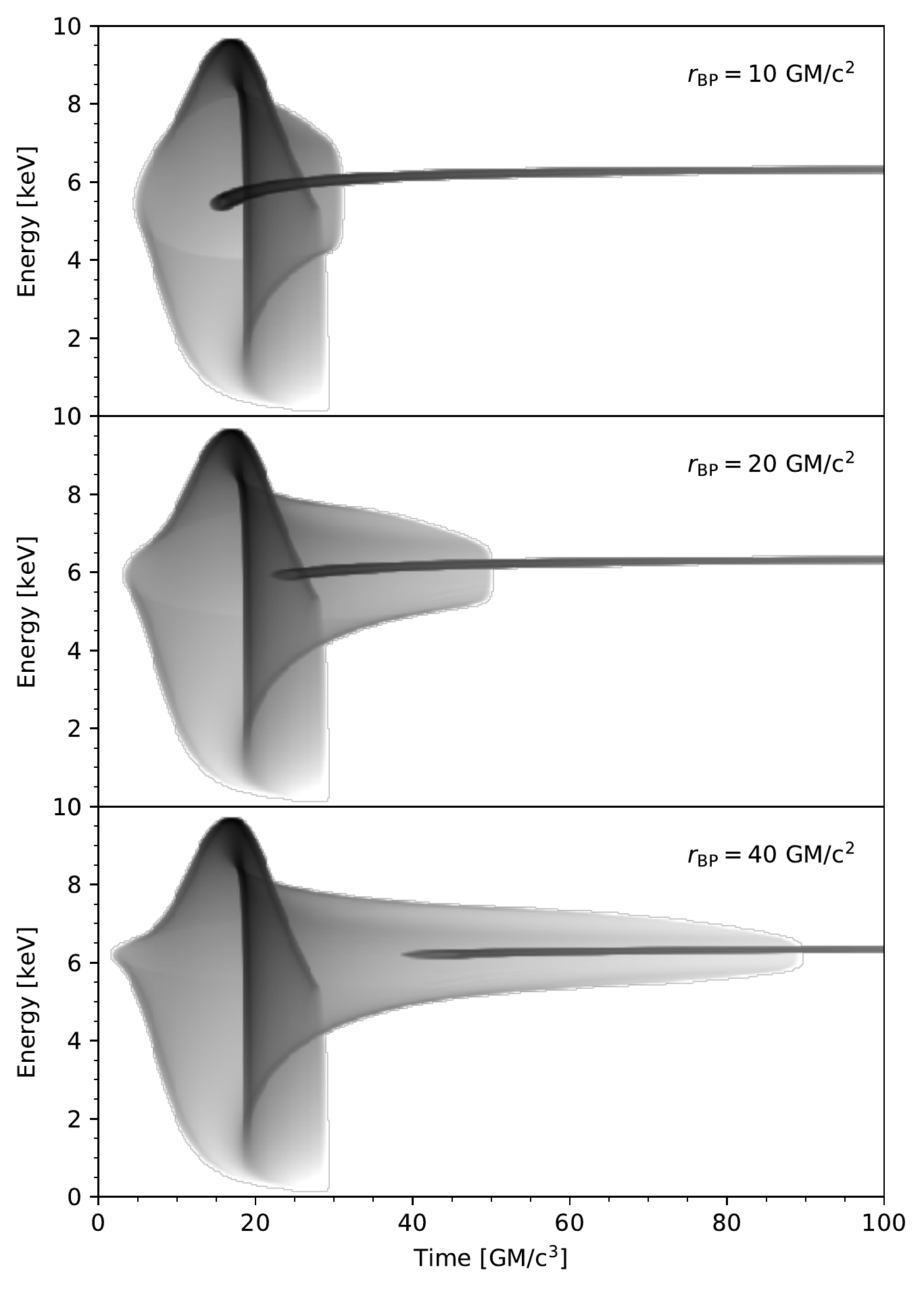}
 \caption{Time evolution of fluorescent iron line spectrum from BP discs with different $r_{\rm BP}$.
 Parameters: black hole spin $a=0.9987$, inner disc inclination $i_{\rm in}=85^\circ$, outer disc inclination $i_{\rm out}=5^\circ$.
 \label{fig:rtr}}
\end{figure*}

\begin{figure}
 \includegraphics[width=\columnwidth]{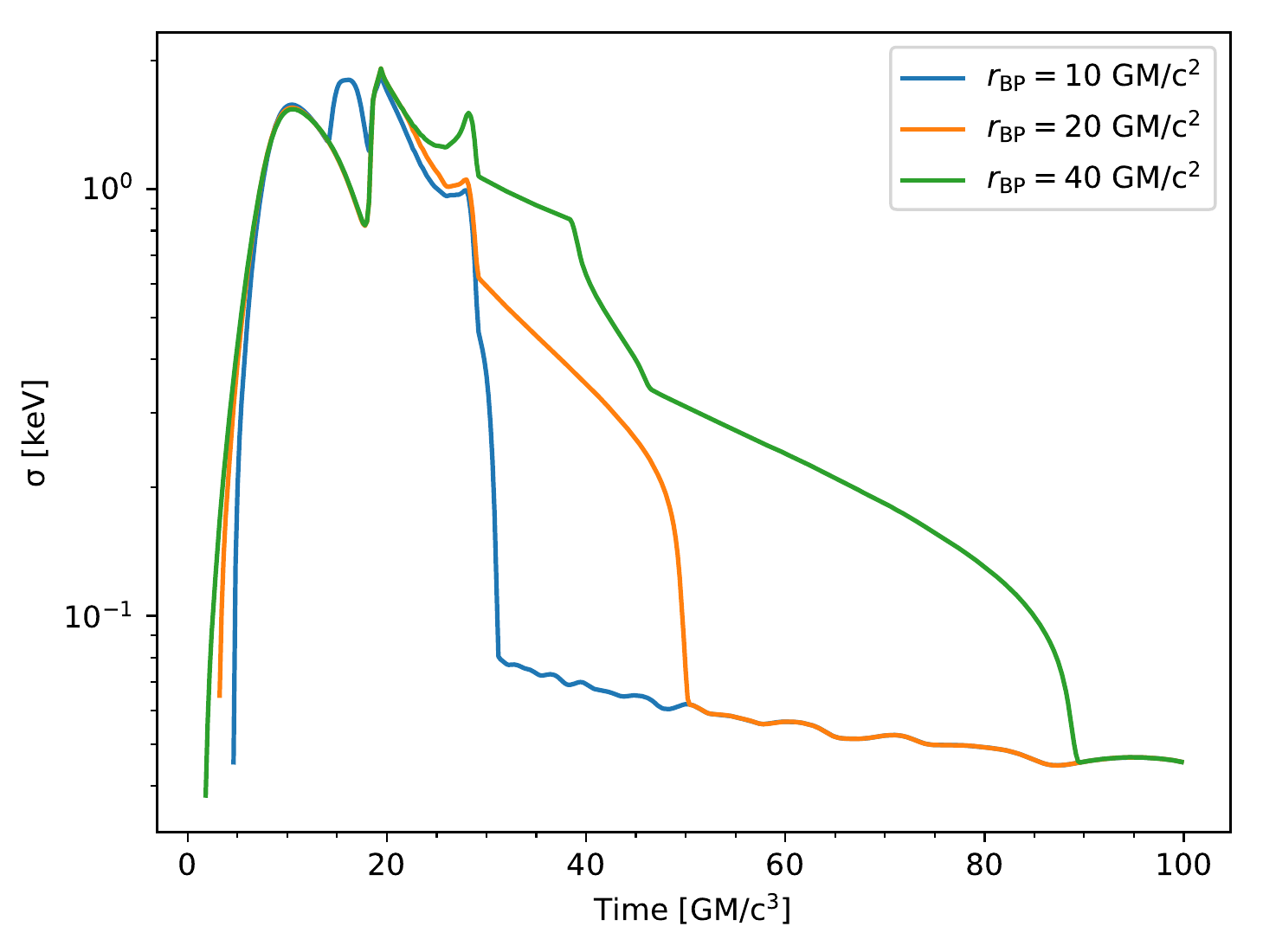}
 \caption{The evolution of the iron line standard deviation for different $r_{\rm BP}$, corresponding to Fig.~\ref{fig:rtr}.\label{fig:rtr_sigma}}
\end{figure}

In Fig.~\ref{fig:rtr} we show the effect of varying the Bardeen-Petterson radius, with $a=0.9987$,
$i_{\rm in}=85^\circ$, and $i_{\rm out}=5^\circ$.
Obviously, the larger $r_{\rm BP}$ is, the later the time-resolved line profile transits from loop shape to tail shape.
We also present time evolution of $\sigma$ in Fig.~\ref{fig:rtr_sigma}, which shows that time for $\sigma$ to 
drop to $\sim0.05~\rm keV$ increase with $r_{\rm BP}$. 
This is expected as the larger $r_{\rm BP}$ is, the later could we see the narrow line from the outer disc under this disc configuration.

\section{Robustness}
\label{sec:robustness}
The emissivity law of the fluorescent line still bears uncertainty.
For the lamp-post geometry, a ``standard'' value of $q=3$ is expected at large radius in the Newtonian case. However, 
in the general relativistic case, for a compact X-ray emitting corona located close to the black hole,
steeper emissivity is predicted for the innermost part of the disc mainly due to the light-bending effect. 
The value of $q$ is not fixed, but depends on various parameters, including the properties
of the corona \citep{wilkins_understanding_2012,gonzalez_probing_2017}, the ionization state of the accretion
\citep{svoboda_origin_2012}, and the angular directionality of the iron line emission \citep{svoboda_role_2009}.

\begin{figure}
 \plotone{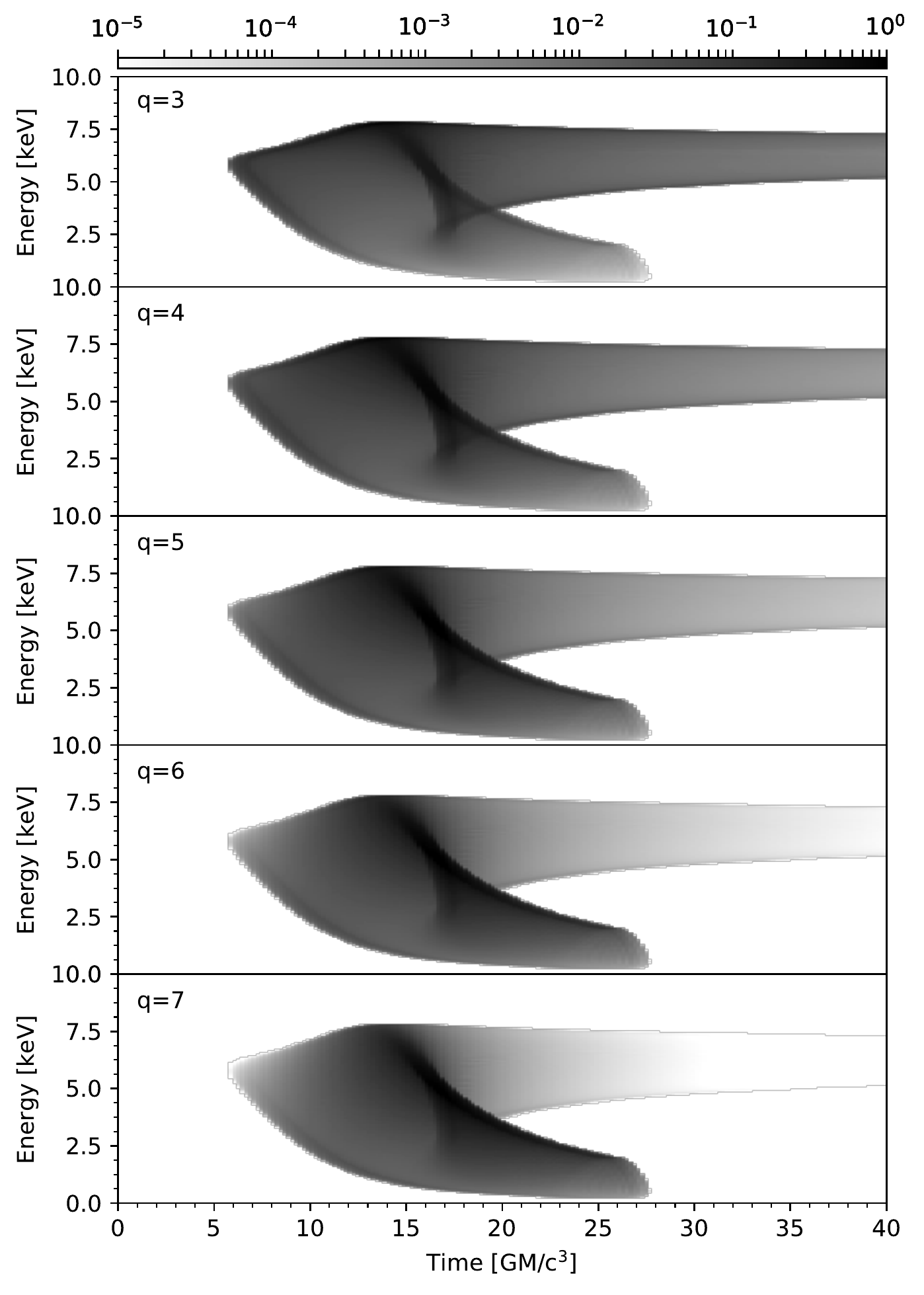}
 \caption{Time-resolved iron line spectra from aligned discs with an inclination of 
 $60^\circ$ around black holes with spin of 0.9987. We assume different emissivity indices $q$ for the fluorescent
 iron line, as indicated in the upper-left corner of each panel.\label{fig:index}}
\end{figure}

In this work we are more interested in the geometry of the
misaligned disc. Therefore in the calculations we take $q=3$ for simplicity. To see whether our main conclusion still holds under steeper emissivity,
we calculate the time-resolved iron line spectra for aligned discs with
inclination of $60^\circ$ while assuming various values of $q$. The black hole is a maximally spinning black hole. 
In the top panel of Fig.~\ref{fig:index} we present the spectrum taking $q=3$, while the spectra assuming steeper emissivity are shown in lower panels.
It is apparent that our particular choice of $q$ does not affect the shape of the ``loop'' structure, the key feature
to constrain the black hole spin and disc inclination.

\section{\textit{eXTP}/LAD simulation}
\textit{The enhanced X-ray Timing and Polarimetry mission} \citep[\textit{eXTP,}][]{zhang_enhanced_2019} is a Chinese X-ray mission
with strong European participation that is
currently in phase B. If approved, it is planned to be launched in 2026.
The Large Area Detector\footnote{http://www.isdc.unige.ch/extp/the-extp-payload.html}(LAD) aboard \textit{eXTP} is an 
ideal instrument for performing time-resolved
spectroscopy, given its large effective area (3.4 $m^2$ between 6 and 10 keV)
and moderate energy resolution (better than 250 keV at 6 keV).

In the simulation, we assume that the variability of the continuum is the same with a relativistic TDE \swsixteen{} during
its \textit{XMM-Newton} observation taken on April 14, 2011 (obsID: 0678380101).
We process the pn data with \textsc{epproc} in \textsc{SAS 18.0.0}, and extract the time-averaged energy spectrum and source light curve
following \citet{kara_relativistic_2016}.
The time resolution of the lightcurve is taken to be the $50~\rm s$,
approximately the time for light to travel $1$ gravitational radius of a $10^7~\rm M_\odot$ black hole.

We fit the time-averaged spectrum with \textsc{Xspec 12.10.1}. The continnum model is a powerlaw continuum
absorbed by interstellar media in both the Milky Way and its host galaxy. We also add 
a Gaussian component to account for the emission line detected by \citet{kara_relativistic_2016}.
The time-averaged $1-10~\rm keV$ flux of the powerlaw component was $8.06\times 10^{-11} ~\rm erg~s^{-1}cm^{-2}$.
Optical spectroscopic observation indicated that the host galaxy of \swsixteen{} is located at a redshift 
of $z=0.3534$ \citep{levan_extremely_2011}.
We calculate the luminosity distance taking the Planck 2015 cosmological parameters \citep{planck_collaboration_planck_2016}
and derive the time-averaged luminosity of the powerlaw continuum.

In the simulations we take the same continuum model. Given the redshift of the source to be simulated,
we calculate the luminosity distance and then derive the time-averaged flux of the powerlaw component with the luminosity obtained above.
For the powerlaw continuum we assume that only the normalization of the powerlaw is
varying with time while the photon index is kept constant.
We take the time resolution of the simulation to be $50~\rm s$, the same with the lightcurve.
For each time bin we simulate the spectrum while the spectral model consists of the continuum plus a emission line component. 
We calculate the normalization of the powerlaw component in the continuum with the time-averaged flux, the count rate of the corresponding
time bin in the lightcurve, and the mean count rate of the lightcurve.
The time-resolved emission line component is calculated by convoluting the powerlaw continuum variability with the response.
The response of the line is the same with what is shown in Fig.~\ref{fig:case0}, except that the neutral iron K$\alpha$ line is redshifted
to a lower energy.
The amplitude of the emission line is set in such a way that the equivalent width is $60~\rm eV$, the equivalent width of the line detected in the
\textit{XMM-Newton} observation \citep{kara_relativistic_2016}. 
We run the same simulation for multiple times if we need a long exposure time.
The simulation is done with
\textsc{Sherpa} in \textsc{CIAO} 4.11\footnote{http://cxc.harvard.edu/ciao/}. 
To perform the simulation, we use the redistribution matrix file LAD\_40mod\_200eV.rmf, 
auxiliary response file LAD\_40mod\_200eV.arf, and background file LAD\_40mod\_200eV.bkg.

\subsection{A relativistic TDE at low redshift}

\begin{figure}
 \includegraphics[width=\columnwidth]{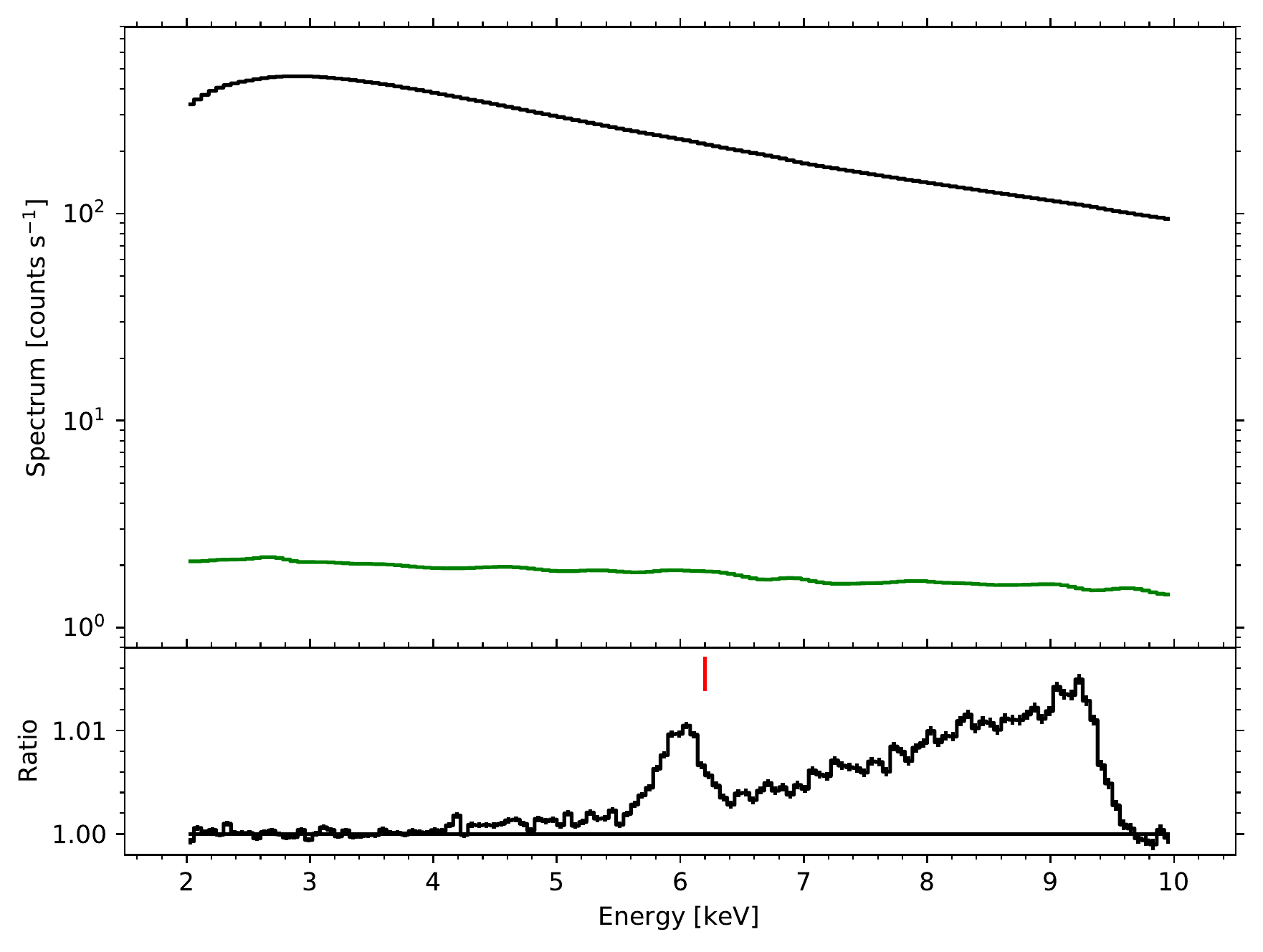}
\caption{Upper panel: the background-subtracted, time-averaged spectrum in black and the background spectrum in green, for
a relativistic TDE at low redshift extracted from a simulated \textit{eXTP}/LAD observation with total exposure time of $35~\rm ks$.
Lower panel: the data-to-continuum ratio.
The vertical red bar indicates the rest-frame energy of the neutral iron K$\alpha$ line in the observer's frame.\label{fig:ladave_tde}}
\end{figure}
\begin{figure}
 \includegraphics[width=\columnwidth]{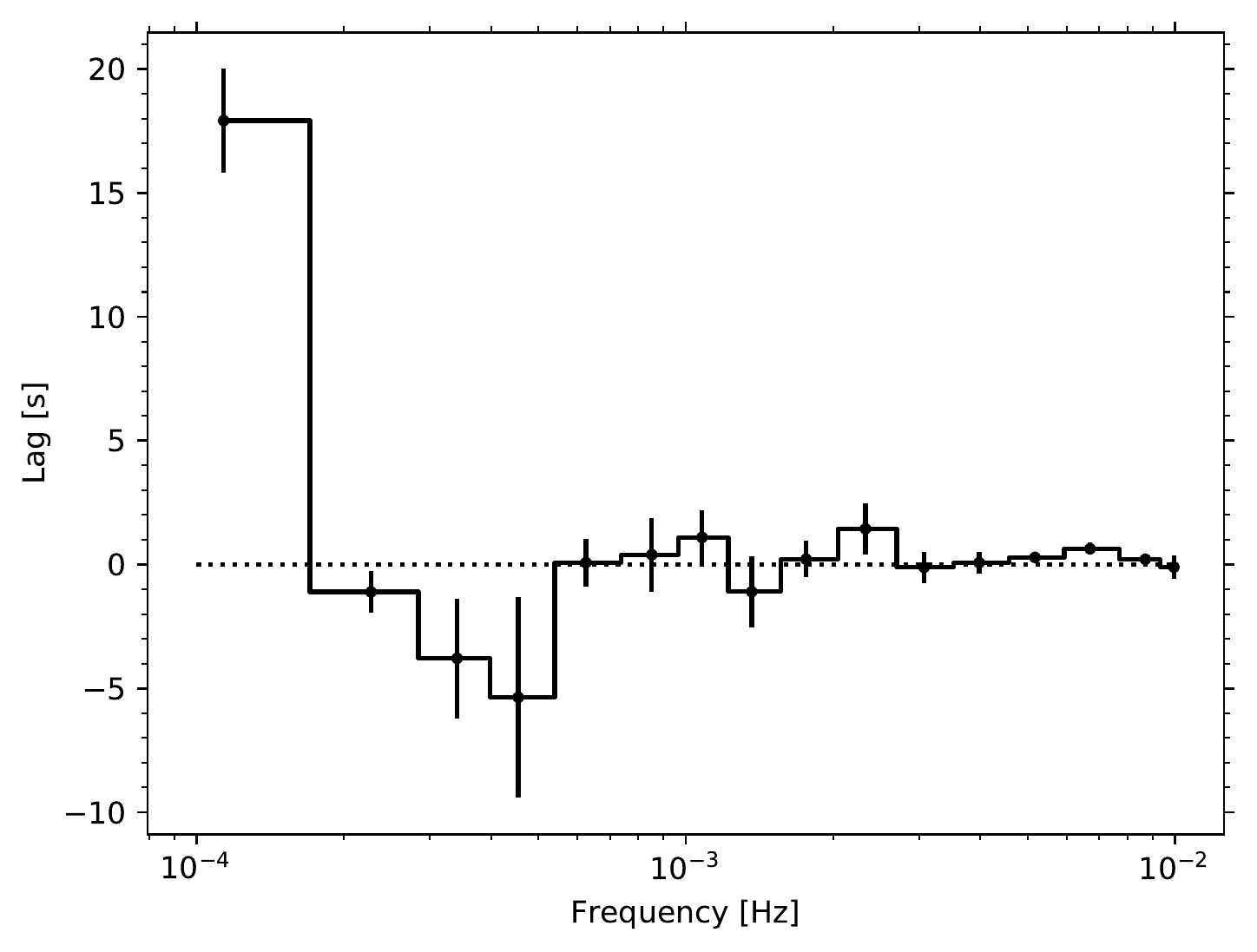}
\caption{The frequency-resolved time lag of the $5.75-6.25~\rm keV$ band versus the $2-4~\rm keV$ reference band, for a relativistic
TDE at low redshift. The lightcurves of the two bands are extracted from a simulated \textit{eXTP}/LAD observation
with a total exposure time of $35 ~\rm ks$.\label{fig:lagspec}}
\end{figure}

We start with a relativistic TDE located at a much lower redshift than
\swsixteen{}. We take the redshift to be 0.032, the mean
redshift of ASAS-SN TDEs \citep{holoien_six_2016,kochanek_tidal_2016}.
In Fig.~\ref{fig:ladave_tde} we present the time-averaged, background subtracted spectrum and the data-to-continuum
ratio for a simulation with an exposure of $35~\rm ks$. The ratio reveals that the time-averaged emission line consists of one narrower red
component peaking at $\sim 6$ and one broader blue component peaking at $\sim 9$ keV.
We take $2-4$ keV as the reference band and produce the frequency-dependent time lag of the $5.75-6.25$ keV band with
respect to the reference band, following the standard procedure \citep[e.g.,][]{nowak_rossi_1999,uttley_x-ray_2014}.
This band contains the centroid of the red component of the emission line, as seen in Fig.~\ref{fig:ladave_tde}.
The time-lag is presented in Fig.~\ref{fig:lagspec}. The longest time lag is $\sim 20$ s, seen at $1.14\times10^{-4}~\rm Hz$. 
Towards high frequency, the time lag drops rapidly to zero.

\begin{figure}
 \includegraphics[width=\columnwidth]{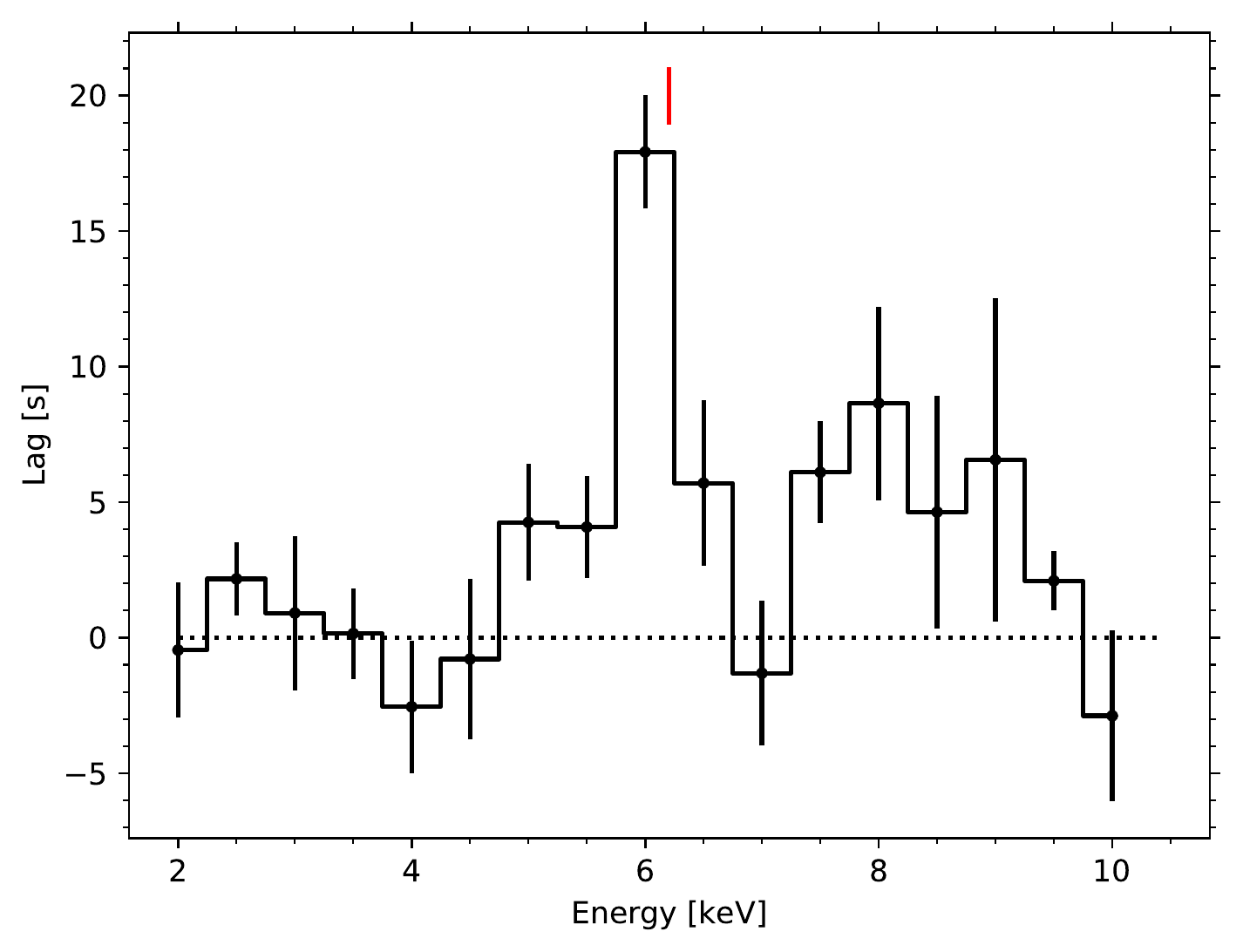}
\caption{The time lag at $1.14\times10^{-4}~\rm Hz$ of various energy bands versus the $2-4 ~\rm keV$ reference band,
for a relativistic TDE at low redshift. The lightcurves of the different bands are extracted from a simulated \textit{eXTP}/LAD observation
with total exposure time of $35~\rm ks$.
\label{fig:lag_tde}}
\end{figure}

In Fig.~\ref{fig:lag_tde} we present the time lag at $1.14\times10^{-4}~\rm Hz$ as a function of energy. 
The energy bands containing the emission line are lagging the reference band, as expected. 
The time lags are quite short as they are highly diluted, given that the emission line contributes a small fraction to the total flux, 
even at the centroids of the red and blue components, as seen in Fig.~\ref{fig:ladave_tde}.
It's worth noting that the time lag of the red component ($\sim 6$ keV)
is longer that that of the blue component ($7-9~\rm keV$).
Given the smaller data-to-continuum ratio of the red component compared with the blue component (the lower panel of Fig.~\ref{fig:ladave_tde}),
the time lag of the red component is more diluted than that of the blue one, indicating that the contrast of the intrinsic time lag between the red and blue
components is even larger than observed.
A longer time delay from the red spectral line component is expected from the BP disc as the red component 
originates from the outer disc with longer time delay.
This demonstrate that we are able to probe the geometry of the BP disc with time-resolved spectroscopy.

\subsection{\swsixteen{}}
\begin{figure}
 \includegraphics[width=\columnwidth]{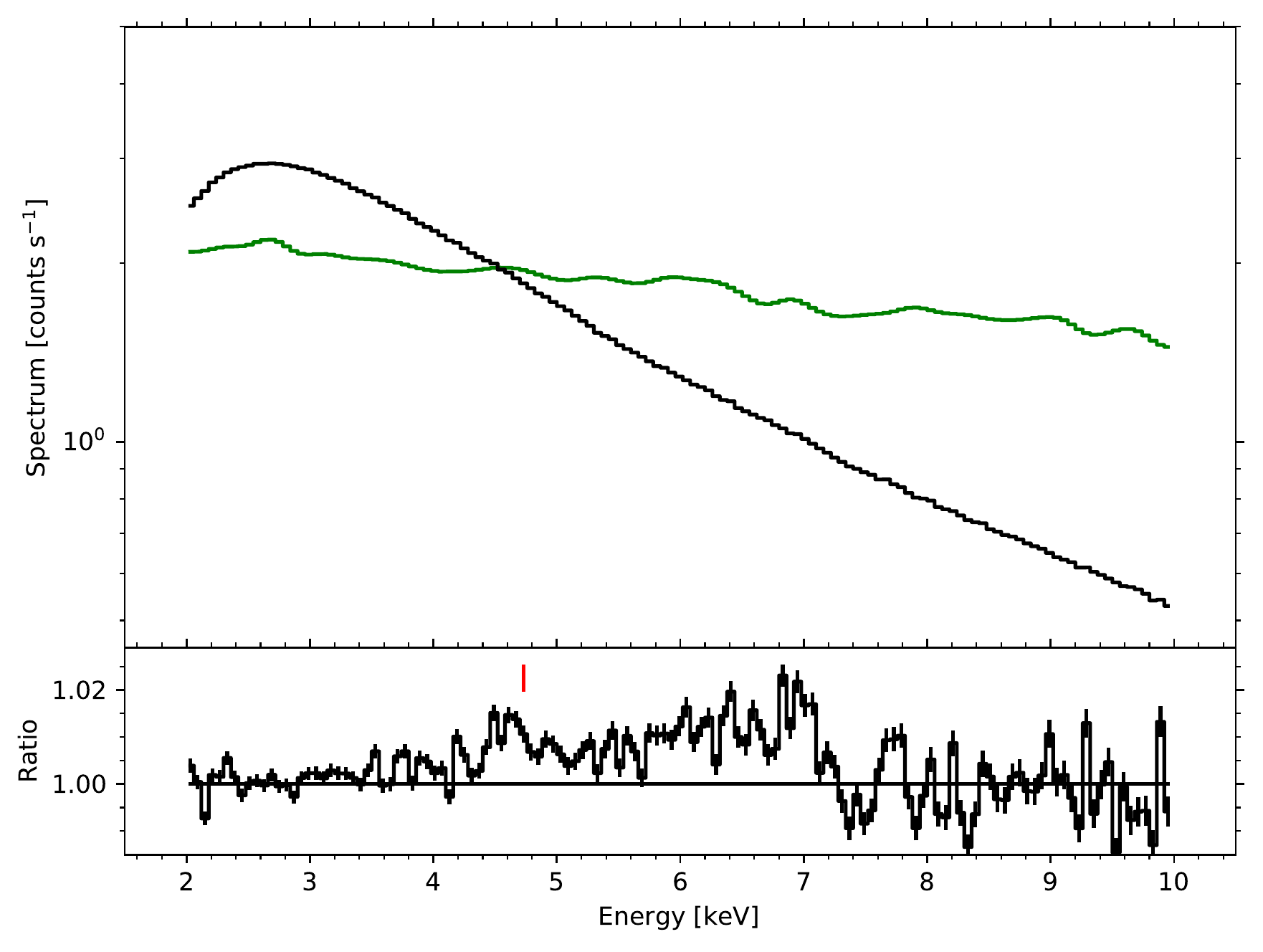}
\caption{The same with Fig.~\ref{fig:ladave_tde}, but for \swsixteen{} with an exposure time of $88~\rm ks$.\label{fig:ladave_1644}}
\end{figure}

\begin{figure}
 \includegraphics[width=\columnwidth]{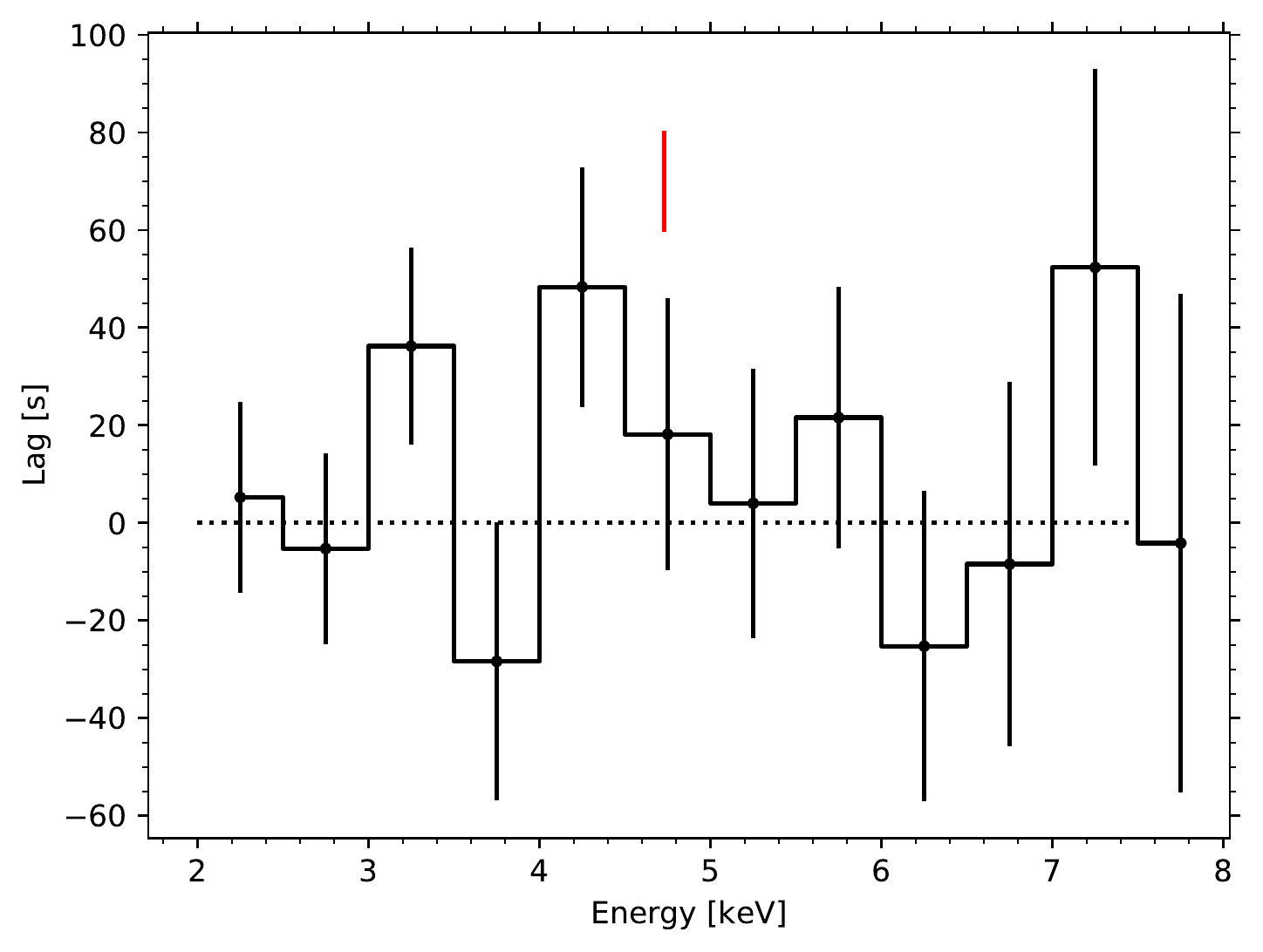}
\caption{The same with Fig.~\ref{fig:lag_tde}, but for \swsixteen{} with an exposure time of $88~\rm ks$.\label{fig:lag_1644}}
\end{figure}

We perform the simulation for \swsixteen{}, following the same procedure. Given its higher redshift (z=0.3534), 
its flux is much smaller. To obtain a better signal-to-noise ratio, we take the exposure time to be $88~\rm ks$.
The rest-frame neutral iron K$\alpha$ is redshifted to $4.73~\rm keV$ in the observer's frame.
In Fig.~\ref{fig:ladave_1644} we present the time-average energy spectrum of \swsixteen{}
and find the spectrum to be dominated by the background above $\sim 4.5~\rm keV$. In Fig.~\ref{fig:lag_1644}
we present the energy-resolved time lag at $1.14\times10^{-4}~\rm Hz$. Due to the weak flux of the source,
the time lag is quite noisy and no time lag more significant than $2\sigma$ can be found. However, it is
worth noting that the \textit{XMM-Newton} observation was taken $\sim 14$ days after the BAT detection when the source flux has dropped by 
a factor of $\sim35$ from the peak flux.

\section{Summary}
We investigate the evolution of the fluorescent iron line spectrum originating from BP discs that are
illuminated by a {\it rising} bright hard X-ray TDE flare located at the center of the disc.
For the example we take, in which the disc is composed of an edge-on inner disc and a face-on outer disc,
the time-resolved iron line spectrum shows distinct feature before and after $t\sim50~\rm GM/c^3$, from
loop structure at the earlier stage to tail shape structure. As we showed
in Paper I, the two structures in the energy vs. time plot are typical of time-resolved iron line originating from edge-on and face-on discs.
This demonstrates that the line reverberation method can be a powerful tool to detect BP discs in TDEs.
It is worth noting that our results of spectral line reverberation for BP disks 
in TDEs also apply to spectral lines other than the iron lines.

We further examine the method by varying several parameters on the time-resolved spectrum.
As the effects of varying black hole spin
and inner disc inclination have been extensively studied in Paper I, we pay special attention on
the BP disc $r_{\rm BP}$ and the outer disc inclination $i_{\rm out}$.
As expected, the time-resolved spectrum is quite
sensitive to these two parameters. The larger the outer disc inclination, the broader the iron line in the late stage of the 
TDE flare.
The location of the Bardeen-Petterson radius $r_{\rm BP}$ affects the time at which the time-resolved line profile transits.
This demonstrates that by application of iron line reverberation we will be able to detect the BP effect,
measure the instantaneous BP radius and put constraint on the black hole mass and spin. 

One of the parameters in our calculation, the emissivity index of the fluorescent iron line $q$, still bears a large range of uncertainty. 
We compare the time-resolved iron line spectra assuming different values of $q$, and find that
the shape of the spectra are nearly the same.

We also simulate time-resolved \textit{eXTP}/LAD spectra of several TDEs.
For a relativistic TDE at a low redshift, 
with a $35~\rm ks$ observation, we would be able to detect distinct features of BP disc with time-resolved spectroscopy.


\acknowledgments
We thank the anonymous referee for his/her careful reading of the manuscript and useful comments and suggestions.
We acknowledge Czech Ministry of Education grant LTI17018 that supports international collaboration in relativstic astrophysics.
WY was supported in part by the National Program on Key Research and Development Project (Grant No. 2016YFA0400804) 
and by the National Natural Science Foundation of China under grant No. 11333005 and U1838203.
WY also would like to acknowledge FAST fellowship supported by Special Funding for Advanced Users, 
budgeted and administrated by Center for Astronomical Mega-Science, Chinese Academy of Sciences (CAMS).
V.K. acknowledges the Czech Science Foundation grant No. 19-01137J.

\end{document}